\documentclass[DIV=calc, paper=letter, fontsize=11pt]{scrartcl}	 
\usepackage[english]{babel} 
\usepackage[protrusion=true,expansion=true]{microtype} 
\usepackage{amsmath,amssymb,amsfonts,amsthm} 
\usepackage[svgnames, table]{xcolor} 
\usepackage[hang, small,labelfont=bf,up,textfont=it,up]{caption} 
\usepackage{booktabs} 
\usepackage{fix-cm}	 
\usepackage[]{units}

\usepackage{tabularx} 

\usepackage[colorlinks=true, linkcolor=blue, citecolor=blue, filecolor=blue, runcolor=blue, urlcolor = blue]{hyperref}

\usepackage{fullpage}
\usepackage{sectsty} 

\usepackage{fancyhdr} 
\usepackage{lastpage} 

\usepackage{nicefrac}
\usepackage{cite}

\usepackage{tikz}
\usetikzlibrary{calc,patterns,decorations.pathmorphing,decorations.markings}
\usepackage{pgfplots}

\usepackage{multicol}
\usepackage{float} 

\newcounter{tempcolnum}
\makeatletter
\newcommand{\multicolinterrupt}[1]{
\setcounter{tempcolnum}{\col@number}
\end{multicols}
#1%
\begin{multicols}{\value{tempcolnum}}
}
\makeatother

\theoremstyle{definition}
\newtheorem{definition}{Definition}[section]

\usepackage{graphicx}
\DeclareGraphicsExtensions{.png}

\usepackage{lettrine} 
\newcommand{\initial}[1]{ 
\lettrine[lines=3,lhang=0.3,nindent=0em]{
\color{DarkGoldenrod}
{\textsf{#1}}}{}}

\usepackage{caption}

\usepackage{graphicx}

\usepackage{titling} 

\newcommand{\HorRule}{\color{DarkGoldenrod} \rule{\linewidth}{1pt}} 

\pretitle{\vspace{-80pt} \begin{flushleft} \HorRule \fontsize{18}{18} \color{DarkRed} \selectfont} 

\title{A topological study of protein folding kinetics}

\posttitle{\par\end{flushleft}} 
\preauthor{\vspace{-5pt} \begin{flushleft}\fontsize{14}{14} \large  \color{DarkRed}} 
\author{Eleni Panagiotou$^{\#}$ $^{\dag}$ and Kevin W. Plaxco$^{\S}$ $^{\P}$ \\} 
\postauthor{\fontsize{12}{12} \small \color{Black} 
$^{\#}$Department of Mathematics, University of Tennessee at Chattanooga, TN 37403 \\
$^{\dag}$SimCenter, University of Tennessee at Chattanooga, TN 37403 \\
$^{\S}$Department of Chemistry and Biochemistry, University of California Santa Barbara, CA 93106-3080 \\
$^{\P}$  Center for Bioengineering, University of California Santa Barbara, CA 93106-3080 
\vskip 1.5em

\end{flushleft} \vspace{-18pt} \HorRule} 

\definecolor{issuePJA_color}{rgb}{1.0,0.0,0.0}

\definecolor{commentPJA_color}{rgb}{1.0,0.0,0.8}

\definecolor{commentEP_color}{rgb}{1.0,0.0,0.8}

\usepackage{soul}

\date{}

\DeclareGraphicsExtensions{.png}

\begin{document}

\maketitle 

\thispagestyle{fancy} 


\initial{F}\textbf{ocusing on a small set of proteins that i) fold in a concerted, “all-or-none” (two-state) fashion and ii) do not contain knots or slipknots, we show that the Gauss linking integral, the torsion and the number of sequence-distant contacts provide information regarding the folding rate. Our results suggest that the global topology/geometry of the proteins shifts from right-handed to left-handed with decreasing folding rate, and that this topological change is associated with an increase in the number of more sequence-distant contacts.}

\section{Introduction}

Proteins, a diverse set of macromolecules each comprised by a unique sequence of amino acids, are a fundamental building block of living organisms \cite{Alberts2002}. In order for proteins to function they must attain conformations that belong in a small subset of the possible 3-dimensional conformations a macromolecule can attain, called the \textit{native state}. \textit{Protein folding} is the process by which a protein molecule folds into this unique, three-dimensional conformation. 
Understanding protein folding is important in predicting of the native structure of a protein from its amino acid sequence which could lead to a better understanding of protein function as well as to the engineering of biopolymers with desired functions. In this study we use topological tools to understand how the topology of the protein relates to its folding rate, the rate with which it spontaneously folds from a random state to the native state.

The 3-dimensional conformation (also known as \textit{tertiary structure}) is determined by the protein sequence (called \textit{primary structure}) \cite{Lawrence2010}. Some patterns of 3-dimensional structure that appear often in proteins are $\alpha$-helices and $\beta$-strands. The sequence of such patterns is called the \textit{secondary structure} of the protein. Structure is a very important clue to understanding and manipulating biological function. However, protein folding is very complex and it has been very difficult to provide a model for it. An interesting fact is that the rate at which proteins fold varies significantly from one protein to the other. At first, this might seem intuitive, since the length of the proteins also varies and one might expect a longer macromolecule to require a longer time to organize itself into a particular 3-dimensional structure. However, studies have shown that the protein length correlates only weakly with the protein folding rate \cite{Plaxco2000}. This lead  to the study of the topological complexity of the native state\cite{Micheletti2003,Plaxco1998,Plaxco2000,Makarov2003,Maxwell2005,Broom2015}. In \cite{Plaxco1998}, a measure of topological complexity, the \textit{relative contact order}, was introduced and showed a very strong correlation with protein folding rates. Moreover, a simpler measure, the \textit{number of sequence distant contacts}, which depicts the sequence distant amino acids that are close in 3-space, provided a similarly strong correlation with the protein folding rate. The number of sequence distant contacts is a global measure of structure which captures features beyond those captured in secondary structure motifs. Its strongly statistically significant correlation with experimental folding rates highlights the extent to which the topology of the native state influences the folding process.
The number of contacts was also successfully used to provide a model for protein folding \cite{Makarov2003}. The underlying idea is that a high number of sequence distant contacts corresponds to few local interactions, suggesting that the route from the unfolded conformation to the native state is slow due to the need to overcome the barriers arising from these spatial restraints. However, it is unknown if the number of sequence distant contacts is a proxy for some more physically meaningful topological measure, which could help us improve our understanding of protein folding kinetics.

In this study we explore the ability of various parameters more directly associated with topology to predict protein folding rates. 
To do so we have looked at the sequence of CA atoms (all amino acids possess a central, or $\alpha$, carbon) of the proteins as vertices connected by edges, such that we can represent the protein as a polygonal curve. To explore the complexity of such curves we invoke knot theory. Specifically, a measure of entanglement that can be directly applied to open chains and is a continuous function of the chain coordinates is the Gauss linking integral. This measure has been successfully applied to polymers in order to study polymer entanglement and it has been shown that it correlates with the physical properties of polymeric material \cite{Panagiotou2010,Panagiotou2013b,Panagiotou2014,Panagiotou2017}. The Gauss linking integral has likewise been used to classify protein conformations successfully \cite{Arteca2000,Rogen2003} and it was shown that the maximum linking between any two parts of a protein correlates with its folding rate \cite{Baiesi2017}. In this study we further use the Gauss linking integral to reveal new characteristics of protein entanglement relevant to protein folding kinetics that can be useful in future models of protein folding.
More precisely, we study the correlation between folding rate and the geometry/topology of the entire protein and also that of the secondary structures that compose it.

The paper is organized as follows: in Section \ref{sec:measures} we define the measures of geometrical/topological complexity we use to characterize protein structure and Section \ref{sec:results} presents our results.

\section{Characterization of protein structure}\label{sec:measures}

\subsection{The number of contacts}

A way in which a straight conformation of a chain differs from a more complex, curved conformation is that in the straight conformation, the 3-space distance of the $i$-th and $j$-th monomer of the molecule is $|i-j|$ times the bond length. This is not the case for any other conformation, where chains whose sequential distance is $|i-j|$ may have been brought closer in 3-space.
Therefore, a way to measure the topological complexity of the native state of a protein is to measure how much this property deviates from that of the straight configuration, by accounting for the number of sequence distant contacts \cite{Plaxco1998}.

We say that two monomers form a contact if their CA-CA distance is less than 6 {\AA} (a value typical for amino acids in close physical contact in the native structure) and their sequence distance greater than 12 amino acids. Closer contacting pairs were excluded because contacts of closer amino acids are trivial as these are enforced by the connectivity of the chain. We denote by $Q_D$ the number of sequence distant contacts in a protein.
It has been shown that the number of contacts and measures related to that (the absolute and relative contact order) correlate well with the protein folding rate \cite{Plaxco1998,Broom2015}.
Moreover, in \cite{Makarov2003} the number of sequence-distant contacts was used to provide a model for protein folding.
In this paper we will not provide a model for protein folding, but we will examine the correlations of several topological parameters with folding rates and their relation to the number of contacts.

\subsection{Gauss linking number and the Total Torsion}

A measure of the degree to which polymer chains interwind and attain complex configurations is given by the Gauss linking integral:

\begin{definition}\label{lk} (Gauss Linking Number). The Gauss \textit{Linking Number} of two disjoint (closed or open) oriented curves  $l_1$ and $l_2$, whose arc-length parametrizations are $\gamma_1(t),\gamma_2(s)$ respectively, is defined as the following double integral over $l_1$ and $l_2$ \cite{Gauss1877}:

\begin{equation}\label{Gausslk}
L(l_1,l_2)=\frac{1}{4\pi}\int_{[0,1]}\int_{[0,1]}\frac{(\dot\gamma_1(t),\dot\gamma_2(s),\gamma_1(t)-\gamma_2(s))}{||\gamma_1(t)-\gamma_2(s)||^3}dt ds,
\end{equation}

\noindent where $(\dot\gamma_1(t),\dot\gamma_2(s),\gamma_1(t)-\gamma_2(s))$
is the \textit{scalar triple product} of $\dot\gamma_1(t),\dot\gamma_2(s)$ and
$\gamma_1(t)-\gamma_2(s)$.
\end{definition}
The Gauss Linking Number is a topological invariant for closed chains and a continuous function of the chain coordinates for open chains.

We also define a one chain measure for the degree of intertwining of the chain around itself:
\begin{definition}\label{Wr_def} (Writhe).  For a curve $\ell$ with arc-length parameterization $\gamma(t)$ is the double integral over $l$:
\begin{equation}\label{Gausslk}
Wr(l)=\frac{1}{4\pi}\int_{[0,1]}\int_{[0,1]}\frac{(\dot\gamma(t),\dot\gamma(s),\gamma(t)-\gamma(s))}{||\gamma(t)-\gamma(s)||^3}dt ds.
\end{equation}
\end{definition}

The Writhe is a continuous function of the chain coordinates for both open and closed chains.
The Average Crossing Number (ACN) is obtained when we consider the absolute value of the integrand in the Writhe. ACN measures the expected number of crossings of a chain with itself in a random projection, while the writhe gives the expected number of signed crossings, which accounts for which strands come over and under. Similarly, the linking number of two chains, measures the expected number of signed inter-crossings of two arcs with each other accounting for which comes over and under in a random projection, divided by two.

The total torsion of a chain, describes how much it deviates from being planar and is defined as:

\begin{definition}\label{torsion} The \textit{torsion} of a curve $\ell$ with arc-length parameterization $\gamma(t)$ is the double integral over $l$:

\begin{equation}\label{tor}
T(l)=\frac{1}{2\pi}\int_{[0,1]}\frac{(\gamma'(t)\times\gamma''(t))\cdot\gamma'''(t)}{||\gamma'(t)\times\gamma''(t)||^2}dt .
\end{equation}

\end{definition}

 Notice that it is possible to construct a random walk with high writhe or torsion that does not contain a knot, however, the mean absolute value of the writhe and torsion increases with knot complexity  \cite{Diao2005,Diao2010,Arsuaga2007a,Panagiotou2010,Millett2004}. In the same way, one can construct a configuration of a non-flat, polygonal chain with zero writhe. However, the writhe of a random walk is non-zero with probability one.

\section{Results}\label{sec:results}

We analyze a set of simple, single domain, non-disulfide-bonded proteins that have been reported to fold in a concerted, “all-or-none”, two-state fashion. Despite their simplicity compared to other proteins, the slowest folding rate of the proteins in the set is a million times slower than the fastest folding rate in the same set. Therefore, this provides a dataset where geometrical and topological features important in protein folding may become aparent. The set of proteins and their corresponding extrapolated folding rates in water can be found in \cite{Plaxco2000}.


\subsection{The number of contacts}

\begin{figure}
\centering
\includegraphics[width=0.75\textwidth]{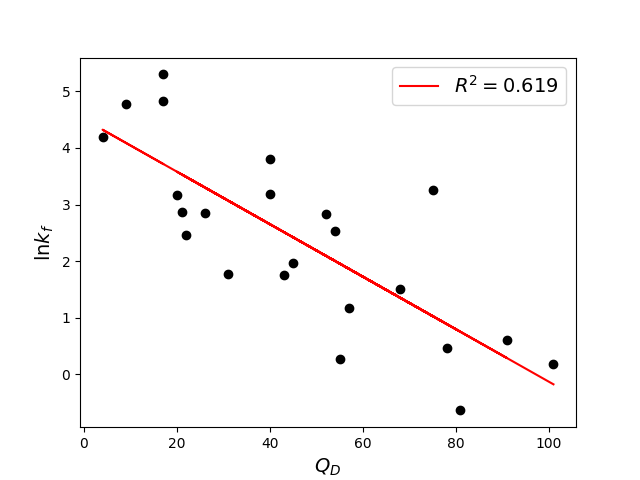}
\caption{The folding rate is decreasing with increasing number of sequence distant contacts.}
\label{fig:QD}	
\end{figure}

Figure \ref{fig:QD} shows the folding rate as a function of the number of sequence-distant contacts, $Q_D$, in a protein. We see that the number of contacts increases with decreasing folding rate. This suggests that the more global interactions are required to form the native state, the slower folding is. The correlation coefficient for the relationship between $\log k_f$ and $Q_D$ is $R^2=0.619$.

\subsection{The writhe and total torsion of the proteins}

\begin{figure}
\centering
\includegraphics[width=0.45\textwidth]{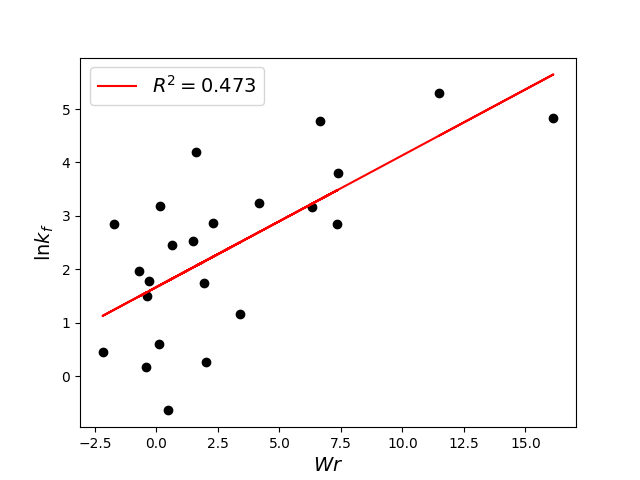}
\includegraphics[width=0.45\textwidth]{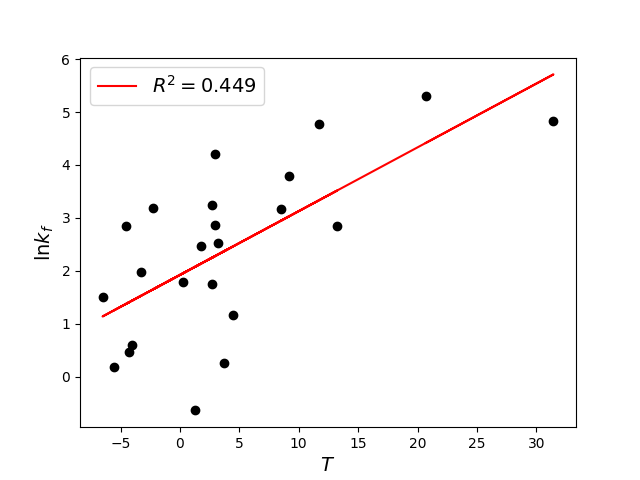}
\caption{The writhe and torsion of the proteins. It is known that the folding rate is influenced by the number of helices. A decreasing number of helices may lead to a decrease of the writhe and torsion of the proteins.}
\label{fig:wr}	
\end{figure}

The set of proteins we analyze in this study do not contain knots or slipknots and their length is in the range of 50-150 amino acids. The mean absolute writhe of unknotted random coils of lengths in the range of the proteins under study is small and one might expect the writhe of the proteins to be small as well \cite{Diao2010,Diao2003,Panagiotou2010,Panagiotou2014}. 
Figure \ref{fig:wr} shows the folding rate as a function of the writhe and torsion of the proteins. The values of the writhe of the proteins significantly exceed the expected values of comparable length random walks. The writhe and the torsion take mostly positive values and are decreasing with decreasing folding rate with correlation coefficients $R^2=0.473$ and $R^2=0.449$, respectively.

At first the fact that the folding rate decreases with decreasing writhe and torsion of the proteins might seem counterintuitive, since one might expect folding rates to decrease as topological complexity increases. However, we notice that the writhe is influenced by the presence of secondary structures. The helices found in proteins are almost invariably right handed due to steric (physical excluded volume) interactions between individual amino acids (which are themselves chiral) and the chirality of the helix. This causes them to contribute a large, positive amount of writhe. $\beta$-strands have particular configurations that have a small in absolute value, but negative writhe (see Section \ref{sec:secstruct}). It is also known that the folding rate decreases with decreasing number of helices, while the number of $\beta$-strands increases \cite{Malik2017}. Therefore, the observed behavior of the writhe and torsion of the proteins may be reflecting exactly this change in secondary structure. Notice that this change is local and may be hiding the underlying global topology/geometry of the protein. In order to extract information about the global geometry of a protein we propose to study the conformation of its \textit{primitive path}. We define the primitive path (PP) of the protein (inspired by the tube model for polymer melts \cite{Rubinstein2003}) to be the axis of the thinnest  tube that surrounds the chain, with no self-intersections and whose diameter (approximately 6 {\AA}) is that of a helix. 

To compute the writhe of the PP in practice, we obtain a semi-analytical formula for the writhe of the PP of the protein using the Gauss linking integral only, without constructing the tube.

First, we notice that the writhe of the protein can be expressed as:

\begin{eqnarray}\label{eq:protwr}
& Wr(\text{protein})=\sum_{i\in S}Wr(s_i)+\sum_{j\in\text{ coils}}Wr(\text{coil } j)+\sum_{i,j \in S} Lk(s_i,s_j)\\ \nonumber
&+\sum_{i \in S}\sum_{j \in \text{coils}}Lk(s_i,\text{coil }j)+\sum_{i,j\in\text{ coils}}Lk(\text{coil } i,\text{coil} j)
\end{eqnarray}

\noindent where $S$ denotes the set of secondary structure elements of the protein.

The protein PP is formed by replacing the protein secondary structure elements by their axis, giving PP$=\lbrace{e_i| e_i \text{ axis of secondary element } i\rbrace}\cup \lbrace \text{coils}\rbrace$. Notice that $Wr(e_i)=0$ for all axes $e_i$. Thus,

\begin{eqnarray}\label{eq:PPwr}
Wr(\text{PP})&=\sum_{i,j\in S'} Lk(e_i,e_j)+\sum_{i, j\in S'}Lk(e_i,\text{coil j})+\sum_{j\in\text{ coils}}Wr(\text{coil }j)\\ \nonumber
&+\sum_{i,j\in\text{ coils}}Lk(\text{coil } i,\text{coil} j)
\end{eqnarray}

\noindent where $S'=\lbrace{e_i| e_i \text{ axis of secondary element } i\rbrace}$.

We notice that, since the secondary structure elements lie in disjoint cells (disjoint convex hulls) from coils and from each other and due to sign cancellations, we can make the following approximations:

\begin{eqnarray}\label{eq:approx1}
Lk(e_i,\text{coil }j)\approx Lk( s_i,\text{coil j})
\end{eqnarray}

and

\begin{eqnarray}\label{eq:approx2}
Lk(e_i, e_j)\approx Lk( s_i,s_j)
\end{eqnarray}

Using Eq. \ref{eq:approx1} and \ref{eq:approx2} in Eq. \ref{eq:PPwr} and comparing it with Eq. \ref{eq:protwr}, the writhe of the PP can be obtained by the writhe of the protein and its secondary structure elements as:

\begin{equation}
Wr(\text{PP})\approx Wr(\text{protein})-\sum_{\alpha-\text{helices}}Wr_{\alpha-\text{helix}}-\sum_{\beta-\text{strands}}Wr_{\beta-\text{strand}}
\end{equation}

Similarly, we define the ACN and the torsion of the PP to be:

\begin{equation}
ACN(\text{PP})=ACN(\text{protein})-\sum_{\alpha-\text{helices}}ACN_{\alpha-\text{helix}}-\sum_{\beta-\text{strands}}ACN_{\beta-\text{strand}}
\end{equation}

and

\begin{equation}
T(\text{PP})= T(\text{protein})-\sum_{\alpha-\text{helices}}T_{\alpha-\text{helix}}-\sum_{\beta-\text{strands}}T_{\beta-\text{strand}}
\end{equation}

\begin{figure}
\centering
\includegraphics[width=0.45\textwidth]{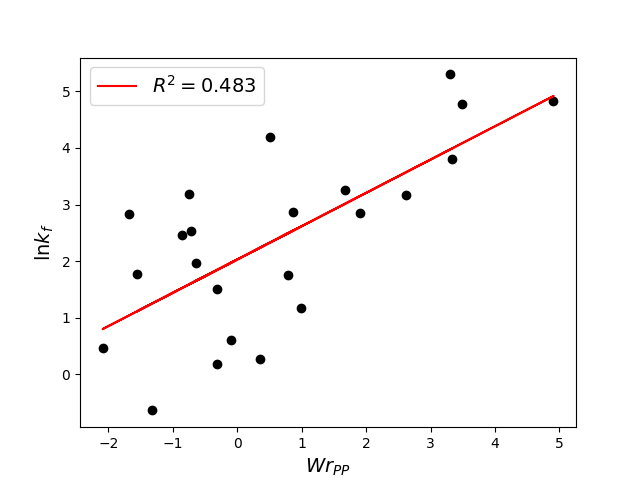}
\includegraphics[width=0.45\textwidth]{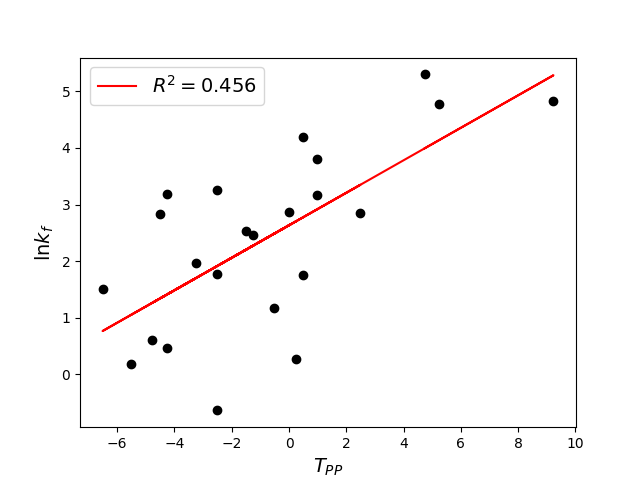}
\caption{Writhe and torsion of the primitive path. The folding rate is decreasing with the writhe and the torsion of the PP shifting from positive to negative values.}
\label{fig:wrpp}	
\end{figure}

The folding rate as a function of the writhe and the torsion of the PP are shown in Figure \ref{fig:wrpp}. The range of values of the writhe and the torsion is now comparable to those of random walks of similar lengths. We see that the folding rate is decreasing with decreasing writhe and torsion of the PP which start from positive values and attain negative values with correlation coefficients $R^2=0.483$ and $R^2=0.456$, respectively. This reveals a significant change in the global conformation of proteins which affects their folding rate: the proteins fold more slowly when they need to attain a configuration with negative global writhe and torsion. This may suggest a role of handedness of proteins not only in local organization but also in global structure.

\begin{figure}
\centering
\includegraphics[width=0.45\textwidth]{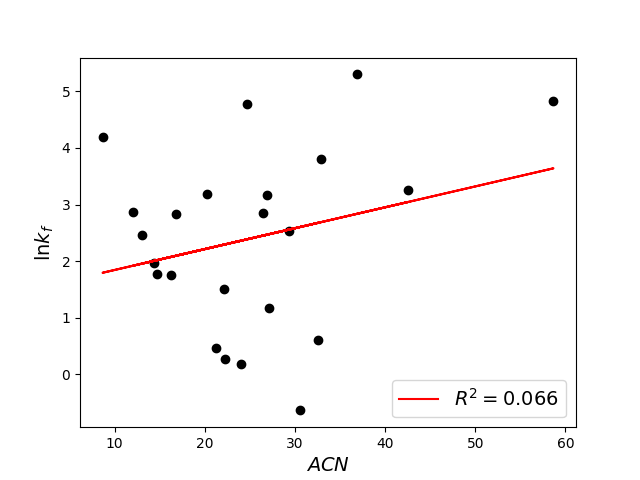}\includegraphics[width=0.45\textwidth]{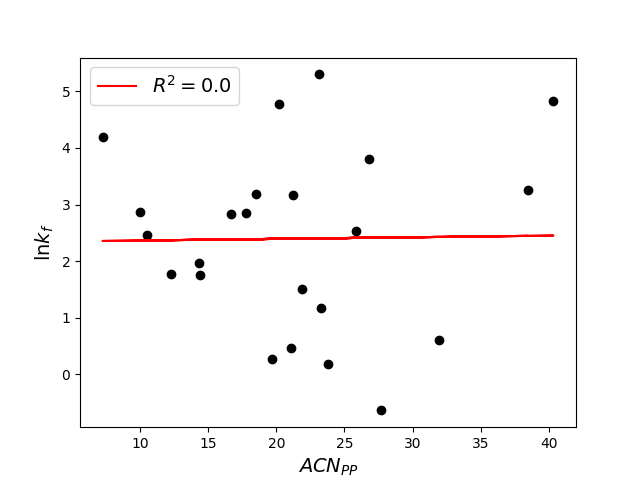}
\caption{The ACN of the protein and its PP. The ACN of the PP shows that the complexity is not decreasing even when the writhe is zero, confirming that a change in global handedness of the protein may be the reason for decreasing folding rates.}
\label{fig:acnpp}	
\end{figure}

The folding rate as a function of the ACN of the proteins and their PP is shown in Figure \ref{fig:acnpp}. The ACN shows a small correlation with folding rate and its values are larger than those for similar length random walks \cite{Diao2003}. This is expected due to the high ACN values of helices. As the folding rate decreases, the  ACN of the PP remains large, even when the writhe is decreasing, confirming that the decrease of the writhe is not reflecting a decrease in complexity but a preference for left-handed global conformations. This may suggest that some left-handed confromations have low energy but they are more difficult to attain \cite{Onuchic1994}.

\subsection{The topology and geometry of secondary structures}\label{sec:secstruct}

It has been shown that proteins whose native state has only $\alpha$-helices fold more rapidly than proteins that also contain $\beta$-strands, while the proteins that only contain $\beta$-strands fold even more slowly \cite{Nguyen2005,Munoz2016}. 
Our results so far suggest that the folding rate is related to the global rearrangement of the secondary structure elements in space and not on the local structure of $\alpha$-helices and $\beta$-strands.
In order to understand the origins of this global rearrangement which results in negative global writhe, in this section, we study how the different secondary structure elements ($\alpha$-helices, $\beta$-strands and coils) link with each other and how this correlates with the folding rate. More precisely, we examine the linking between $\alpha$-helices, coils and $\beta$-strands. (In the following we report only the results that showed the strongest correlations for this dataset.)

\begin{figure}
\centering
\includegraphics[width=0.45\textwidth]{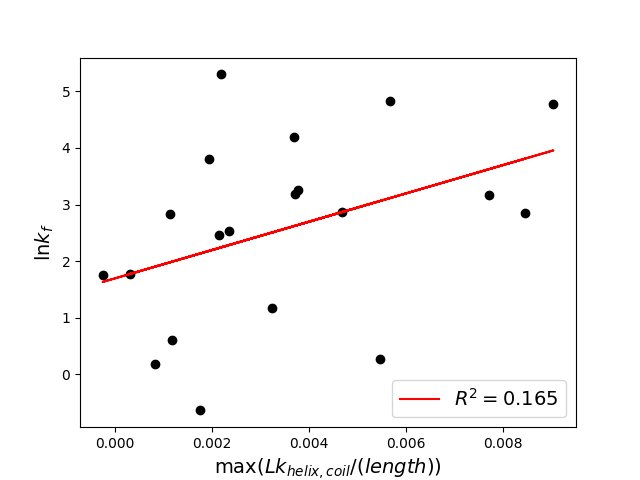}\includegraphics[width=0.45\textwidth]{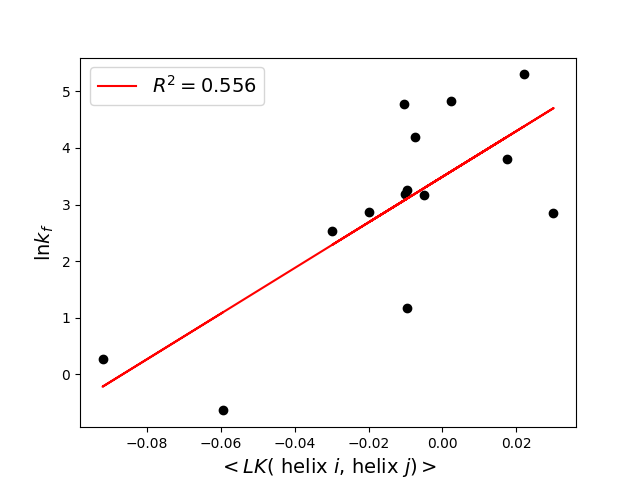}
\caption{Left: The folding rate as a function of the maximum absolute linking number between helices with neighboring coils. Right: The folding rate as a function of the average linking number of helices in a protein. The folding rate is decreasing with decreasing linking number, suggesting that the proteins fold slowly when a left-handed conformation of helices is required. }
\label{fig:lkhccons}	
\end{figure}

When examining the linking of $\alpha$-helices with coils in a protein we notice that the linking number is rather random in sign and we find only a  small correlation with folding rate, $R^2=0.077$. However, we find that the linking of a helix with the neighboring (in secondary structure) coils  is mostly positive and the protein folding rate tends to decrease with decreasing linking. The maximum absolute linking number relative to the length of the coil and helix involved is shown in Figure \ref{fig:lkhccons}. We see that the folding rate is decreasing with decreasing maximum linking number of a helix with a neighboring coil, with $R^2=0.165$. The fact that it is positive shows a right-handed preference for the conformation helix-neighboring coil. The tendency of the protein folding rate to decrease with decreasing maximum linking between helices and coils may indicate an increasing difficulty of the protein to keep right-handed conformations, presumably due (again) to interactions between the chirality of the chain's fold and the chirality of the individual amino acids.
Figure \ref{fig:lkhccons} also shows the average linking number between helices in each protein of the dataset that contains at least two helices. The folding rate decreases as the average linking number between helices  becomes more and more negative, with $R^2=0.556$, confirming our results in the previous section which suggest a change in the preferred handedness of the global linking of the proteins.

\begin{figure}
\centering
\includegraphics[width=0.45\textwidth]{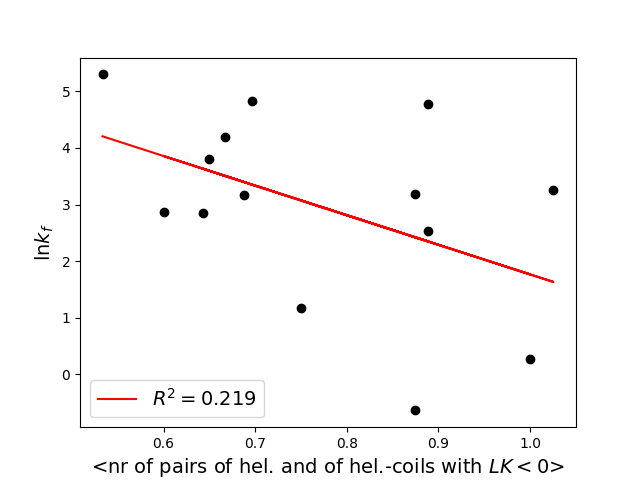}\includegraphics[width=0.45\textwidth]{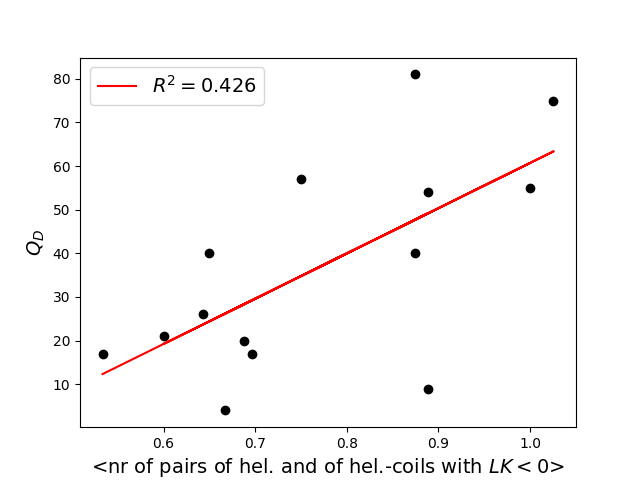}
\caption{The folding rate and the number of sequence distant contacts as a function of the relative number of pairs of $\alpha$-helices or pairs of $\alpha$-helices and coils with negative linking number. }
\label{fig:nhlk}	
\end{figure}

Our data do not show a correlation of the folding rate with the absolute value of the linking between secondary structures, suggesting that the folding rate may depend only on the sign of the linking between secondary structures. To further explore this, we examine if the folding rate correlates with the number of occurrences of pairs of secondary structure elements which have negative linking. 
Figure \ref{fig:nhlk} shows the folding rate as a function of the number of pairs of helices or pairs of helices-coils with negative linking number over the total number of pairs. Figure \ref{fig:nhlk} also shows the number of sequence distant contacts as a function of the number of pairs of helices and helices coils with negative linking. The results suggest that conformations of negative linking helices and conformations of negative linking helices with coils require many contacts, which causes the folding rate to decrease. Conversely, it may be true that negative linking is difficult for the chain to achieve (perhaps due to interactions between the ``handedness" (all amino acids in proteins are of the ``L" chirality) of the chain and the handedness of negative linking interactions, and that the number of sequence distant contacts is a proxy for this property.

\begin{figure}
\centering
\includegraphics[width=0.45\textwidth]{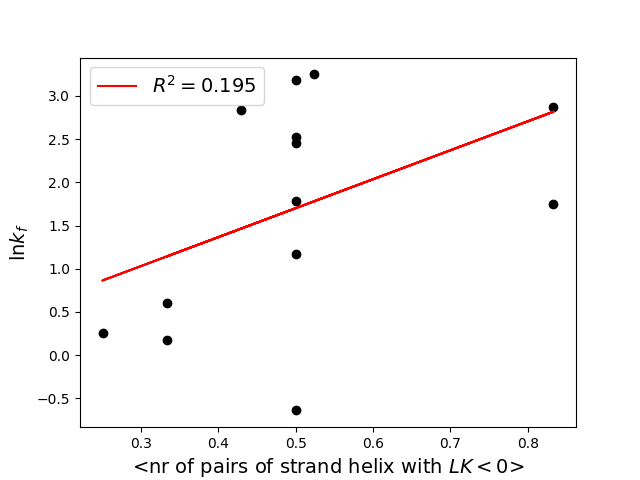}\includegraphics[width=0.45\textwidth]{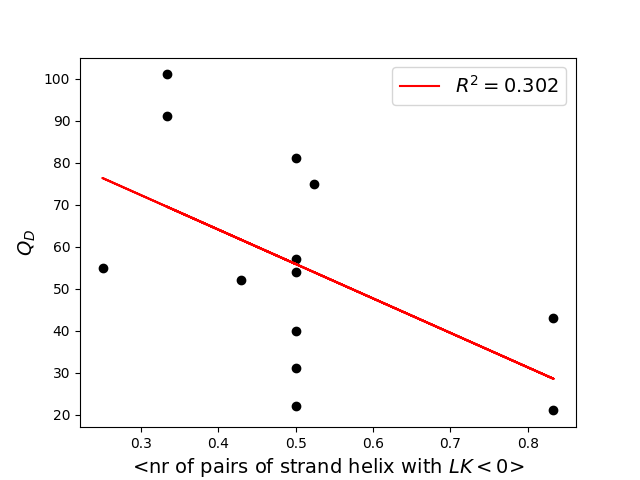}
\caption{The folding rate and the number of sequence distant contacts as a function of the relative number of pairs of $\alpha$-helices and $\beta$-strands with negative linking number. The 0.5 ratios indicate the presence of antiparallel $\beta$-strands which contribute a negative and positive linking number.}
\label{fig:nshlk}	
\end{figure}

The sign of the linking number between $\alpha$-helices and $\beta$-strands is rather random ($R^2=0.089$). However, the folding rate is decreasing with decreasing number of pairs of $\alpha$-helices and $\beta$-strands with negative linking over the total number of pairs, with $R^2=0.195$ (see Figure \ref{fig:nshlk}). We notice a large set of proteins with exactly 0.5 ratio of pairs of helices-strands with negative linking. We believe that this is because of the particular conformations of sheets, with antiparallel $\beta$-strands, which contribute an opposite sign to the average.

\begin{figure}
\centering
\includegraphics[width=0.45\textwidth]{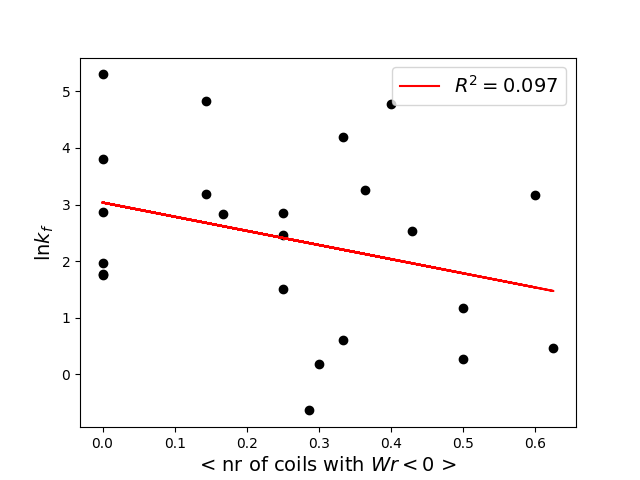}\includegraphics[width=0.45\textwidth]{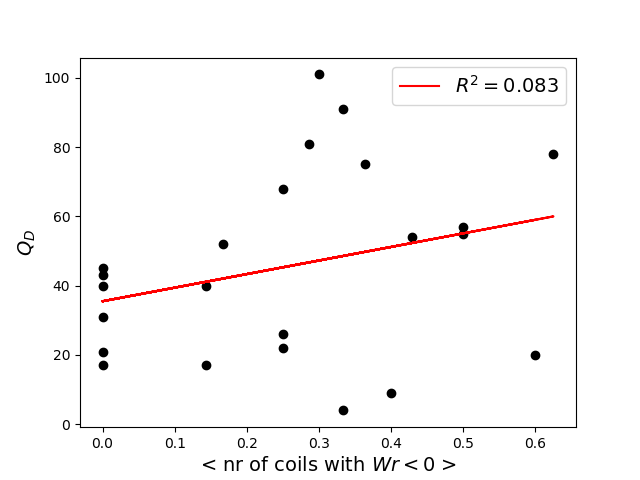}\\
\includegraphics[width=0.45\textwidth]{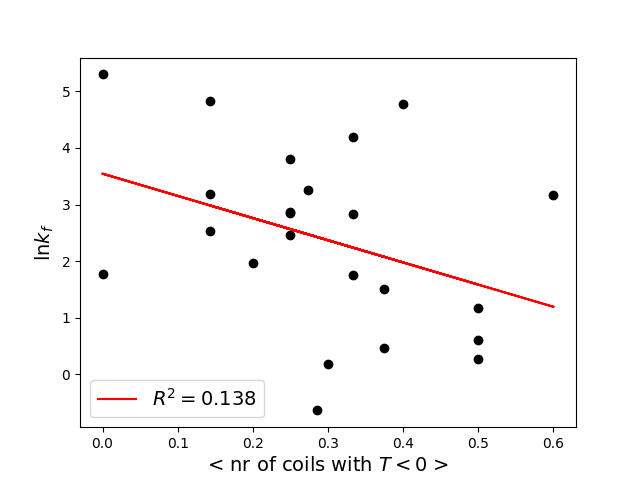}
\includegraphics[width=0.45\textwidth]{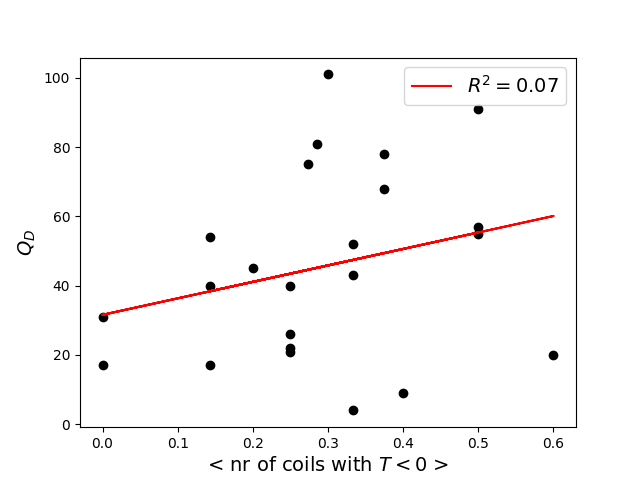}
\caption{The folding rate and the number of sequence distant contacts as a function of the relative number of coils with negative writhe (above) and torsion (below). The folding rate is related to the number of coils with negative torsion, even if their presence does not correlate with the number of sequence distant contacts.}
\label{fig:cwr}	
\end{figure}

These results suggest a dependence of the folding rate simply on the number of negative linking occurrences involving secondary structure elements, and not particularly on the values of these linking occurrences. Since the secondary structure elements follow an almost straight axis, the source of change in their relative orientation stems from the coils in-between. Indeed, the coils can attain any random configuration. Our results show that the writhe of the coils is on average positive, but negative writhe coils also exist and our results suggest that the folding rate decreases as the number of coils with negative writhe is increasing.
In Figure \ref{fig:cwr} we see that folding rate is decreasing with increasing number of coils with negative writhe and also with number of coils with negative torsion over the total number of coils in a protein with $R^2=0.097$ and $R^2=0.138$.

Figure \ref{fig:cwr} shows also the number of sequence distant contacts as a function of the number of coils with negative writhe and as a function of coils with negative torsion. The results suggest that more sequence distant contacts are required for increasing number of negative writhe and negative torsion coils. However, the results show that the folding rate is more sensitive to the number of negative writhe/torsion coils than the number of sequence distant contacts. Therefore, this may be a feature that is only weakly captured by the number of sequence distant contacts but influences the folding rate.

\section{Conclusions}

Studies have shown that the folding rates of proteins that fold in a concerted all-or-none fashion correlate very well with the number of sequentially distant contacts \cite{Makarov2003}. By studying a set of proteins which fold in a concerted, all-or-none fashion and do not contain knots, we showed that their folding rates and the number of sequence-distant contacts correlate with their writhe, ACN and torsion. By examining the global entanglement of the proteins, ignoring the local secondary structure, we showed that the folding rate decreases as the global writhe and torsion of the proteins becomes more and more negative. We find that the folding rate is related to the relative orientation of helices and strands in space, with negative linking conformations being associated with more sequence distant contacts and, in turn, slower folding. We also see that coils with negative writhe or torsion slow down the folding process even if they do not require many contacts.
Our results confirm that the highly organized native structure is too complex to be captured by a single parameter, and suggest that the combination of the topological and geometrical tools, such as the number of sequence distant contacts and the Gauss linking integral, could provide complementary information useful in the search of a model for protein folding kinetics.

\bibliographystyle{plain}
\bibliography{paperDatabase2}{}

\end{document}